\documentclass[twoside,twocolumn,9pt]{article}
\usepackage{extsizes}
\usepackage[super,sort&compress,comma]{natbib} 
\usepackage[version=3]{mhchem}
\usepackage[left=1.5cm, right=1.5cm, top=1.785cm, bottom=2.0cm]{geometry}
\usepackage{balance}
\usepackage{mathptmx}
\usepackage{sectsty}
\usepackage{graphicx} 
\usepackage{lastpage}
\usepackage[format=plain,justification=justified,singlelinecheck=false,font={stretch=1.125,small,sf},labelfont=bf,labelsep=space]{caption}
\usepackage{float}
\usepackage{fancyhdr}
\usepackage{fnpos}
\usepackage[english]{babel}
\addto{\captionsenglish}{%
  
}
\usepackage{array}
\usepackage{droidsans}
\usepackage{charter}
\usepackage[T1]{fontenc}
\usepackage[usenames,dvipsnames]{xcolor}
\usepackage{setspace}
\usepackage[compact]{titlesec}
\usepackage{hyperref}

\usepackage{epstopdf}

\definecolor{cream}{RGB}{222,217,201}

\begin{document}

\pagestyle{fancy}
\thispagestyle{plain}
\fancypagestyle{plain}{
\renewcommand{\headrulewidth}{0pt}
}

\makeFNbottom
\makeatletter
\renewcommand\LARGE{\@setfontsize\LARGE{15pt}{17}}
\renewcommand\Large{\@setfontsize\Large{12pt}{14}}
\renewcommand\large{\@setfontsize\large{10pt}{12}}
\renewcommand\footnotesize{\@setfontsize\footnotesize{7pt}{10}}
\makeatother

\renewcommand{\thefootnote}{\fnsymbol{footnote}}
\renewcommand\footnoterule{\vspace*{1pt}%
\color{cream}\hrule width 3.5in height 0.4pt \color{black}\vspace*{5pt}} 
\setcounter{secnumdepth}{5}

\makeatletter 
\renewcommand\@biblabel[1]{#1}            
\renewcommand\@makefntext[1]%
{\noindent\makebox[0pt][r]{\@thefnmark\,}#1}
\makeatother 
\renewcommand{\figurename}{\small{Fig.}~}
\sectionfont{\sffamily\Large}
\subsectionfont{\normalsize}
\subsubsectionfont{\bf}
\setstretch{1.125} 
\setlength{\skip\footins}{0.8cm}
\setlength{\footnotesep}{0.25cm}
\setlength{\jot}{10pt}
\titlespacing*{\section}{0pt}{4pt}{4pt}
\titlespacing*{\subsection}{0pt}{15pt}{1pt}

\fancyfoot{}
\fancyfoot[LO,RE]{\vspace{-7.1pt}\includegraphics[height=9pt]{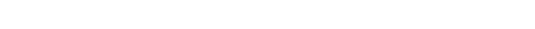}}
\fancyfoot[CO]{\vspace{-7.1pt}\hspace{13.2cm}\includegraphics{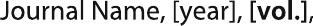}}
\fancyfoot[CE]{\vspace{-7.2pt}\hspace{-14.2cm}\includegraphics{head_foot/RF}}
\fancyfoot[RO]{\footnotesize{\sffamily{1--\pageref{LastPage} ~\textbar  \hspace{2pt}\thepage}}}
\fancyfoot[LE]{\footnotesize{\sffamily{\thepage~\textbar\hspace{3.45cm} 1--\pageref{LastPage}}}}
\fancyhead{}
\renewcommand{\headrulewidth}{0pt} 
\renewcommand{\footrulewidth}{0pt}
\setlength{\arrayrulewidth}{1pt}
\setlength{\columnsep}{6.5mm}
\setlength\bibsep{1pt}

\makeatletter 
\newlength{\figrulesep} 
\setlength{\figrulesep}{0.5\textfloatsep} 

\newcommand{\topfigrule}{\vspace*{-1pt}%
\noindent{\color{cream}\rule[-\figrulesep]{\columnwidth}{1.5pt}} }

\newcommand{\botfigrule}{\vspace*{-2pt}%
\noindent{\color{cream}\rule[\figrulesep]{\columnwidth}{1.5pt}} }

\newcommand{\dblfigrule}{\vspace*{-1pt}%
\noindent{\color{cream}\rule[-\figrulesep]{\textwidth}{1.5pt}} }

\makeatother

\twocolumn[
  \begin{@twocolumnfalse}
{\includegraphics[height=30pt]{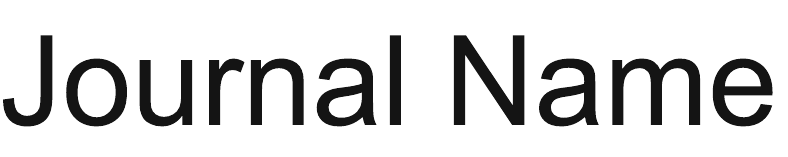}\hfill\raisebox{0pt}[0pt][0pt]{\includegraphics[height=55pt]{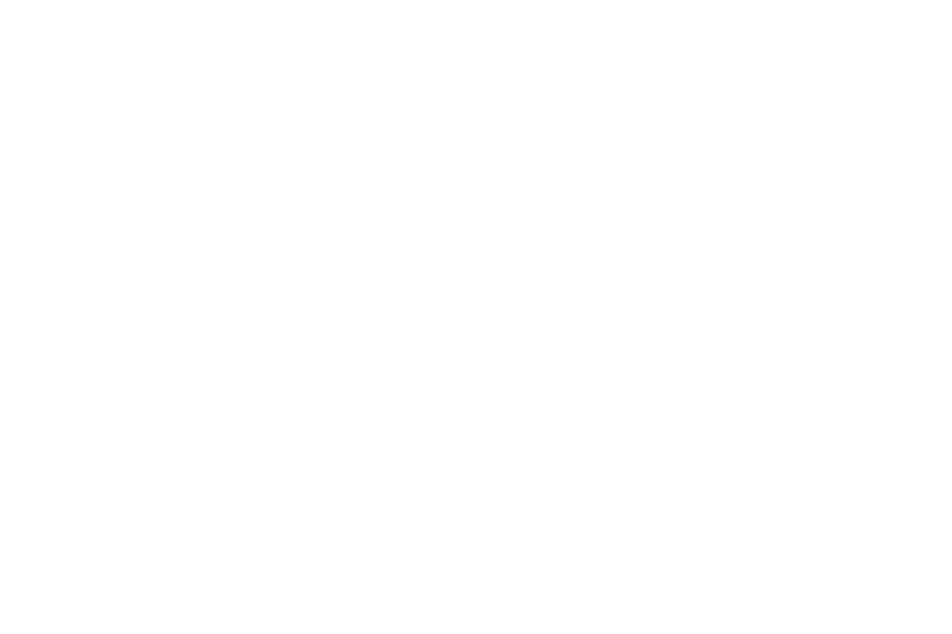}}\\[1ex]
\includegraphics[width=18.5cm]{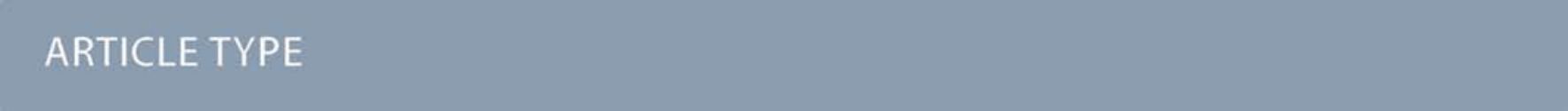}}\par
\vspace{1em}
\sffamily
\begin{tabular}{m{4.5cm} p{13.5cm} }
\includegraphics{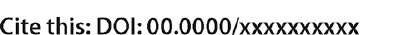} & \noindent\LARGE{\textbf{Enhancing superconductivity in MXenes through hydrogenation$^{\dag}$}} \\
\vspace{0.3cm} & \vspace{0.3cm} \\

 & \noindent\large{Jonas Bekaert,\textit{$^{a,\ast,\ddag}$} Cem Sevik,\textit{$^{a,b,\ddag}$} and Milorad V. Milo\v{s}evi\'{c}\textit{$^{a,\S}$}} \\
 
 \includegraphics{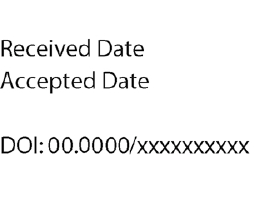} & \noindent\normalsize{Two-dimensional transition metal carbides and nitrides (MXenes) are an emerging class of atomically-thin superconductors, whose characteristics are highly prone to tailoring by surface functionalization. Here we explore the use of hydrogen adatoms to enhance phonon-mediated superconductivity in MXenes, based on first-principles calculations combined with Eliashberg theory. We first demonstrate the stability of three different structural models of hydrogenated Mo- and W-based MXenes. Particularly high critical temperatures of over 30 K are obtained for hydrogenated Mo$_2$N and W$_2$N. Several mechanisms responsible for the enhanced electron-phonon coupling are uncovered, namely (i) hydrogen-induced changes in the phonon spectrum of the host MXene, (ii) emerging hydrogen-based phonon modes, and (iii) charge transfer from hydrogen to the MXene layer, boosting the density of states at the Fermi level. Finally, we demonstrate that hydrogen adatoms are moreover able to induce superconductivity in MXenes that are not superconducting in pristine form, such as Nb$_2$C.} \\

\end{tabular}

 \end{@twocolumnfalse} \vspace{0.6cm}

  ]

\renewcommand*\rmdefault{bch}\normalfont\upshape
\rmfamily
\section*{}
\vspace{-1cm}


\footnotetext{\textit{$^{a}$Department of Physics \& NANOlab Center of Excellence, University of Antwerp, Groenenborgerlaan 171, B-2020 Antwerp, Belgium}}
\footnotetext{\textit{$^{b}$Department of Mechanical Engineering, Faculty of Engineering, Eskisehir Technical University, 26555 Eskisehir, Turkey}}
\footnotetext{\textit{$^{\ast}$ jonas.bekaert@uantwerpen.be}}
\footnotetext{\textit{$^{\S}$ milorad.milosevic@uantwerpen.be}}
\footnotetext{\ddag These authors contributed equally to this work.}

\footnotetext{\dag~Electronic Supplementary Information (ESI) available: Phonon band structures, electronic band structures, Fermi surfaces incl.~Fermi velocities, Eliashberg functions and electron-phonon coupling of all investigated compounds. See DOI: 00.0000/00000000.}



\section{\label{sec:intro}Introduction}

MXenes are one of the most recent additions to the 2D materials family \cite{doi:10.1002/adma.201304138,Naguib2012,Naguib2013}, hosting rich electronic properties, as a result of their diverse charge, orbital and spin degrees of freedom \cite{doi:10.1002/adma.201304138,C7TC00140A,Anasori2017,Gogotsi2019}. In their thinnest form they consist of a layer of carbon or nitrogen sandwiched in between two transition metal layers. Since electronic structure calculations have demonstrated all pristine MXenes to be metallic \cite{C7TC00140A}, they are prime candidates to host superconductivity in the 2D limit. Our recent first-principles exploration has revealed the occurrence of 2D superconductivity in six MXene compounds with superconducting critical temperatures ranging from 2 to 16 K \cite{D0NR03875J}. The highest $T_c$'s were obtained for Mo- and W-based MXenes, with the maximum $T_c$ of 16 K found for Mo$_2$N \cite{D0NR03875J}. In addition, interplay between the superconducting state and a charge density wave phase, manifesting itself as an instability in the phonon dispersion, was found for W$_2$N \cite{D0NR03875J}. 

On the other hand, pristine Nb$_2$C was predicted as non-superconducting \cite{D0NR03875J}, as recently confirmed experimentally \cite{Kamysbayev979}. The same experiment showed, however, that superconductivity can be induced in Nb$_2$C by functionalization with Cl, S, Se, and NH groups, yielding a maximum $T_c$ of 7 K in the latter case \cite{Kamysbayev979}. These results demonstrate the potential of surface-functionalization to induce and enhance superconductivity in MXenes.

In this work, we focus on functionalization of MXenes with hydrogen (H). H has recently been established as a key element to achieve phonon-mediated high-temperature superconductivity. Its inherently high-frequency phonon modes have brought the superconducting critical temperatures ($T_c$) of hydride compounds under high pressure to unrivaled heights, up to room temperature \cite{PhysRevLett.92.187002,Drozdov2019,PhysRevLett.122.027001,Errea2020,Snider2020}. However, the need for ultrahigh pressures -- as high as 267 GPa for the recent carbonaceous sulfur hydride with record-holding $T_c$ \cite{Snider2020} -- severely limits practical applications of this type of room-temperature superconductivity. 

Depositing hydrogen adatoms on a 2D material provides a very effective way to take advantage of the hydrogen-induced boost in the electron-phonon coupling (EPC), without the need for any applied pressure. This principle has been demonstrated for hydrogenated monolayer MgB$_2$ \cite{PhysRevLett.123.077001} -- with a $T_c$ of 67 K, which can be boosted to over 100 K by applying biaxial tensile strain -- as well as for the elemental 2D material gallenene \cite{Petrov_2021}. These systems inherit the stability of the 2D host material, while at the same time they are very rich in hydrogen, as needed for a sizeable effect on the superconducting properties. The enhancement of the EPC originates from (i) hybridization between electronic and vibrational states of hydrogen and the host atoms, and (ii) emerging ultrahigh frequency modes of hydrogen, expanding the frequency range of the EPC \cite{PhysRevLett.123.077001}. In addition to the effect on the superconducting state, our results on stability, structural and electronic properties of hydrogenated MXenes are also expected to stimulate ongoing efforts to employ MXenes for hydrogen storage \cite{Liu2021,KUMAR2021105989}. 

\section{\label{sec:methods}Methodology}

The electron phonon-coupling (EPC) and superconducting properties for all the considered systems are modeled through density functional perturbation theory (DFPT), as implemented within the ABINIT code \cite{Gonze2020}. We used the Perdew–Burke–Ernzerhof (PBE) functional in combination with the relativistic Hartwigsen–Goedecker–Hutter (HGH) pseudopotentials \cite{PhysRevB.58.3641}. Considering the marginal effect of spin-orbit-coupling (SOC) on the superconducting properties of the pristine form of these crystals \cite{D0NR03875J}, we proceeded without taking SOC into account. Within the HGH pseudopotentials, 14 valence electrons for the transition metal elements and 4 (5) valence electrons for C (N) were taken into account. To achieve a high accuracy an energy cutoff of 60 Ha for the plane-wave basis, a dense 32$\times$32$\times$1 \textit{k}-point grid, 16$\times$16$\times$1 \textit{q}-point grid, and at least 16 {\AA} of vacuum in the unit cell were used. Fermi-Dirac smearing with broadening of 0.01 Ha was used throughout the calculations (except where indicated otherwise, namely a reduced value of 0.0025 Ha was used in certain DFPT calculations to identify emerging lattice instabilities). To further characterize the changes in the Fermi surfaces due to hydrogenation we have calculated the Fermi velocities from the electronic band structures by $\mathbf{v}_{F} = \hbar^{-1} \boldsymbol{\nabla}_{\textbf{k}}\epsilon_{\textbf{k}}\mid_{\epsilon=E_{F}}$. 

For the superconducting state we have applied Eliashberg theory, a quantitatively accurate extension to the Bardeen–Cooper–Schrieffer (BCS) theory for phonon-mediated superconductivity \cite{Eliashberg1960,Eliashberg1961,RevModPhys.89.015003}. We have evaluated the spectral function of the EPC, the Eliashberg function, as
\begin{equation*}
    \alpha^{2}F(\omega)=\frac{1}{N_{F}}\sum_{\nu,\mathbf{k},\mathbf{q}}\left| g^{\nu}_{\mathbf{k},\mathbf{k+q}}\right|^{2}\delta(\omega-\omega^{\nu}_{\mathbf{q}})\delta(\epsilon_{\mathbf{k}})\delta(\epsilon_{\mathbf{k+q}})~,
\end{equation*}
using the density of states at the Fermi level ($N_F$), the EPC matrix elements ($g_{\textbf{k},\textbf{k}+\textbf{q}}^{\nu}$), and the phonon ($\omega_{\textbf{q}}^{\nu}$) and electron ($\epsilon_{\textbf{k}}$) band structures (with $E_F$ put to 0) obtained from our \textit{ab initio} calculations. The resulting EPC constant was calculated through $\lambda=2\int_{0}^{\infty}\alpha^{2}F(\omega)\omega^{-1}d\omega$. We have subsequently evaluated the superconducting $T_c$ using the Allen-Dynes formula \cite{PhysRev.167.331,PhysRevB.12.905,ALLEN19831}, and a screened Coulomb repulsion of $\mu^{*}=0.13$ -- the same value as used for the calculation of $T_c$ of the pristine MXenes \cite{D0NR03875J}.

\section{\label{sec:cryst_struct}Crystal structures and stability analysis}

\begin{figure}[t]
\includegraphics[width=\linewidth]{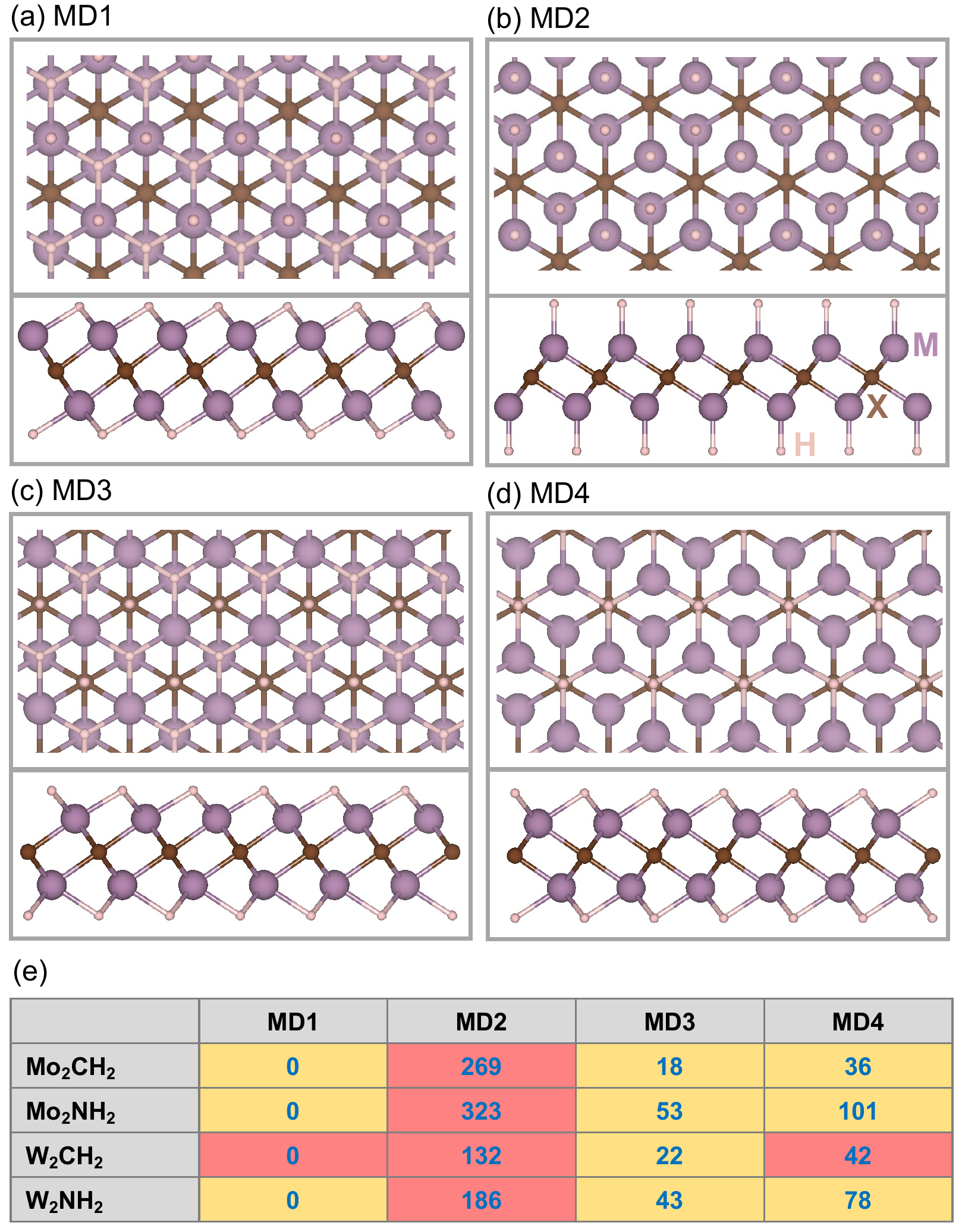}
\caption{\label{figure1} (a)-(d) Four different structural models for H-functionalization of MXenes (named MD 1--4): top and side views. (e) Stability of the different structural models, for each considered M and X atom. Here, yellow and red colors indicate dynamically stable and unstable structures respectively, as obtained from our phonon calculations. Differences in formation energy between the structures w.r.t. MD1 (in meV/atom) are also provided, with lower values indicating higher thermodynamic stability.}
\end{figure}

We considered four different high-symmetry configurations for two-sided hydrogen functionalization, depicted in Fig.~\ref{figure1}(a)-(d). In the first structural model (MD1), the H atoms vertically align with M atoms on the other side of the stack. The second model (MD2) is characterized by direct bonding of H on the M atoms, on both sides. In the third model (MD3), one H atom aligns with the M atom on the other side, while the second H atom aligns with the central X atom. Finally, both H atoms align with the X atom to form the fourth structural model (MD4). 

To assess preferential configurations we carried out an analysis of the dynamical and thermodynamic stability of these structures, as presented in Fig.~\ref{figure1}(e). Certain trends in our stability analysis pertain to all considered MXene compounds. First of all, in all cases MD2 is found to be dynamically unstable. Secondly, the ranking of lowest to highest formation energies of the different structures is the same for all compounds, namely $E_f(\mathrm{MD}1)<E_f(\mathrm{MD}3)<E_f(\mathrm{MD}4)<E_f(\mathrm{MD}2)$. Hence, both from the dynamical and thermodynamic point of view MD2 is unfavorable. The fact that MD1 is identified here as the energetically preferred structure is corroborated by experimental results on Ti$_2$C\textit{T}$_2$ (\textit{T}=S,Cl,Se,Br), for which the same preferred structure has been reported \cite{Kamysbayev979}. W$_2$C stands out from the other compounds in that it is only dynamically stable in MD3 (and even in this case it features strong phonon softening around high-symmetry \textit{k}-point M, as shown in the Supplemental Material). 

The lattice parameters of the host MXenes are to a large extent unchanged by hydrogenation, with all relative changes below $\pm 2\%$. As an example we consider the three cases with the highest $T_c$'s, i.e., Mo$_2$CH$_2$ in MD1, Mo$_2$NH$_2$ in MD3 and W$_2$NH$_2$ in MD1, as will be elaborated on in Section \ref{sec:EPC_SC}. The relative changes in the in-plane lattice parameter respectively amount to $-1.8\%$, $0.3\%$ and $0\%$ and the changes in the M-X bond length $1.9\%$, $1.6\%$ and $0.9\%$. This very limited perturbation of the lattice due to H is in stark contrast with the observation of giant in-plane lattice expansion of Ti$_2$C with Te functionalization ($>18\%$) \cite{Kamysbayev979}. The relaxed M-H bond length is in almost all cases $\sim 2$ \AA, except in the case of the unstable MD2 where it is markedly lower at $\sim 1.7$ \AA. 

\section{\label{sec:elect_vib}Electronic and vibrational properties}

\begin{figure}[t]
\includegraphics[width=\linewidth]{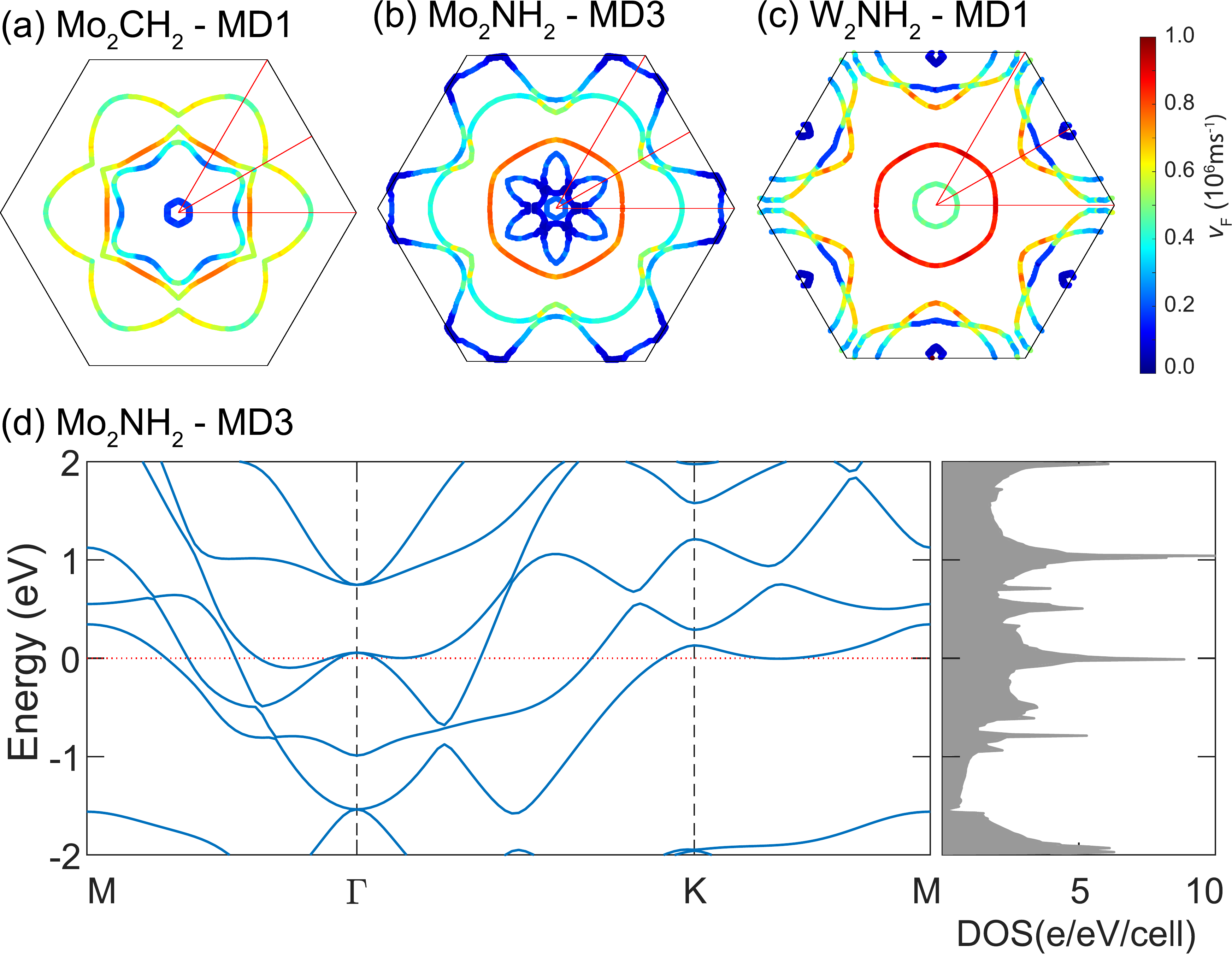}
\caption{\label{figure2} (a)--(c) Fermi surfaces of Mo$_2$CH$_2$ (MD1), Mo$_2$NH$_2$ (MD3) and W$_2$NH$_2$ (MD1) respectively, represented in the $\Gamma$-centered Brillouin zone. The colors indicate the Fermi velocities. (d) Electronic band structure of Mo$_2$NH$_2$ (MD3), with the Fermi level ($E_F$) set to zero, and the corresponding density of states, showing a large peak at $E_F$, boosting the EPC.}
\end{figure}

\begin{figure*}[h]
\includegraphics[width=\linewidth]{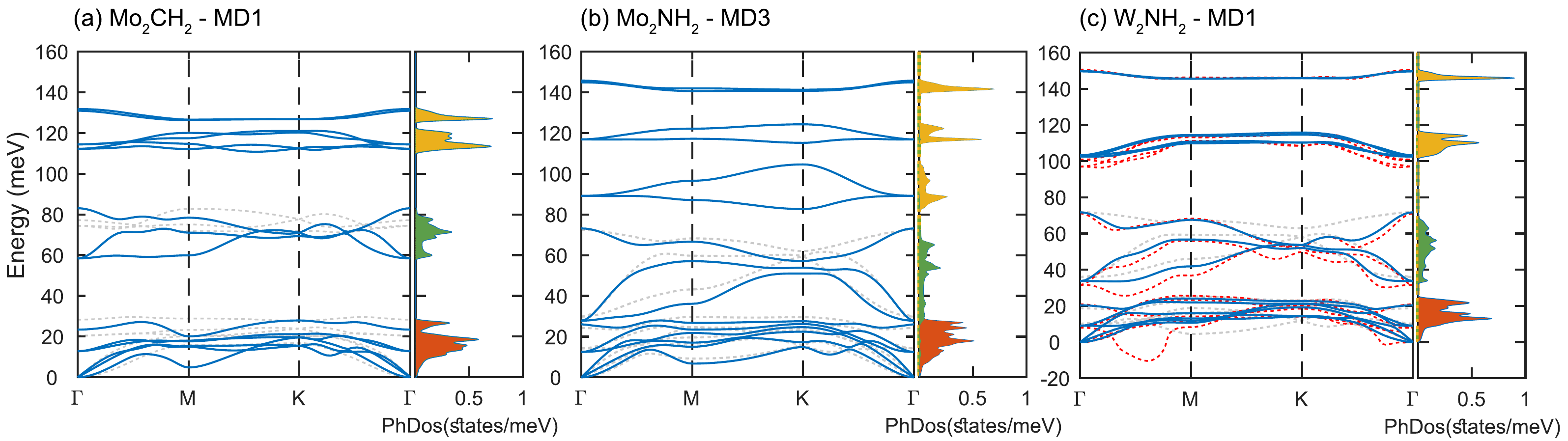}
\caption{\label{figure3} Phonon dispersion and phonon density of states (PhDOS) of (a) Mo$_2$CH$_2$ (MD1), (b) Mo$_2$NH$_2$ (MD3) and (c) W$_2$NH$_2$ (MD1). The grey dashed lines depict the phonon dispersions of the pristine compounds, for comparison. In (c) the dispersion obtained using reduced electronic smearing (0.0025 Ha) is included as a red dashed line, showing a CDW-type instability occurring between points $\Gamma$ and M. The colors in the PhDOS give the contributions of different atomic types: red for the M atom (Mo or W), green for the X atom (C or N) and yellow for H.}
\end{figure*}

Electron doping from hydrogen to the MXene layers induces considerable changes in the electronic band structures and Fermi surfaces. The Fermi surface of pristine Mo$_2$C consists of three distinct groups of sheets. Around $\Gamma$, there are six small electron-like pockets and a larger circular electron-like, while centered around M there is an oval sheet, almost touching the larger $\Gamma$-centered sheet along the $\Gamma$-M direction \cite{D0NR03875J}. In the hydrogenated case, all four Fermi sheets of Mo$_2$CH$_2$ (MD1) are $\Gamma$-centered, and have distinctly six-fold symmetries, as shown in Fig.~\ref{figure2}(a). The loss of Fermi sheets centered around M upon hydrogenation reduces the density of states at the Fermi level ($N_F$) to 2.09 eV$^{-1}$, compared with 3.19 eV$^{-1}$ in the case of pristine Mo$_2$C \cite{D0NR03875J}. Likewise, the Fermi sheets around M have considerably shrunk in MD3 and MD4 of Mo$_2$CH$_2$ (see Supplemental Material), leading to a similar reduction in $N_F$ w.r.t.~the pristine case -- see Table \ref{tab:table1}. 

On the other hand, the DOS at $E_F$ of Mo$_2$N is enhanced upon hydrogenation for all stable structural models w.r.t.~the pristine value of 2.58 eV$^{-1}$ \cite{D0NR03875J}. MD3 of Mo$_2$NH$_2$ is characterized by a particularly high $N_F$ of 6.75 eV$^{-1}$ per unit cell. Its Fermi level coincides exactly with a pronounced peak in the DOS, as shown in Fig.~\ref{figure2}(d). The  corresponding Fermi surface is depicted in Fig.~\ref{figure2}(b), consisting of multiple hole pockets in the vicinity of $\Gamma$, surrounded by a quasi-circular sheet further away from $\Gamma$, and two large entwined sheets wrapping around the Brillouin zone edge. Likewise, the DOS at $E_F$ also increases under the influence of hydrogenation in the case of W$_2$N, compared with the pristine value of 1.96 eV$^{-1}$ per unit cell. While the Fermi surface of pristine W$_2$N only possesses sheets centered around $\Gamma$, multiple sheets appear around M and K upon hydrogenation (see Fig.~\ref{figure2}(c), depicting the case of W$_2$NH$_2$ -- MD1). 

The resulting phonon band structures of selected hydrogenated MXene compounds, obtained from our DFPT calculations, are shown in Fig.~\ref{figure3}, and compared to their pristine counterparts. Overall, in all three compounds, H contributes high-frequency phonon modes in the range of 100--150 meV. In the case of Mo$_2$CH$_2$, the addition of hydrogen also leads to significant softening of the Mo-based ZA phonon mode, particularly around the M point [Fig.~\ref{figure3}(a)]. In this compound, the contributions of the different atoms to the phonon band structure are well-separated, owing to the large differences in atomic masses between Mo, C and H. Akin to the case of pure Mo$_2$N \cite{D0NR03875J}, the phonon modes due to N and H are more dispersive in Mo$_2$NH$_2$ [Fig.~\ref{figure3}(b)]. Pristine W$_2$N was demonstrated to host a charge density wave(CDW)-type instability in its vibrational states, centered around the M point. Our DFPT calculations for W$_2$NH$_2$ in MD1 show that this instability persists, albeit around a different wavevector, namely at 2/3 $\Gamma$-M (Fig.~\ref{figure3}(c)). 

\section{\label{sec:EPC_SC}Enhanced electron-phonon coupling and superconductivity}

\begin{figure}[h]
\includegraphics[width=\linewidth]{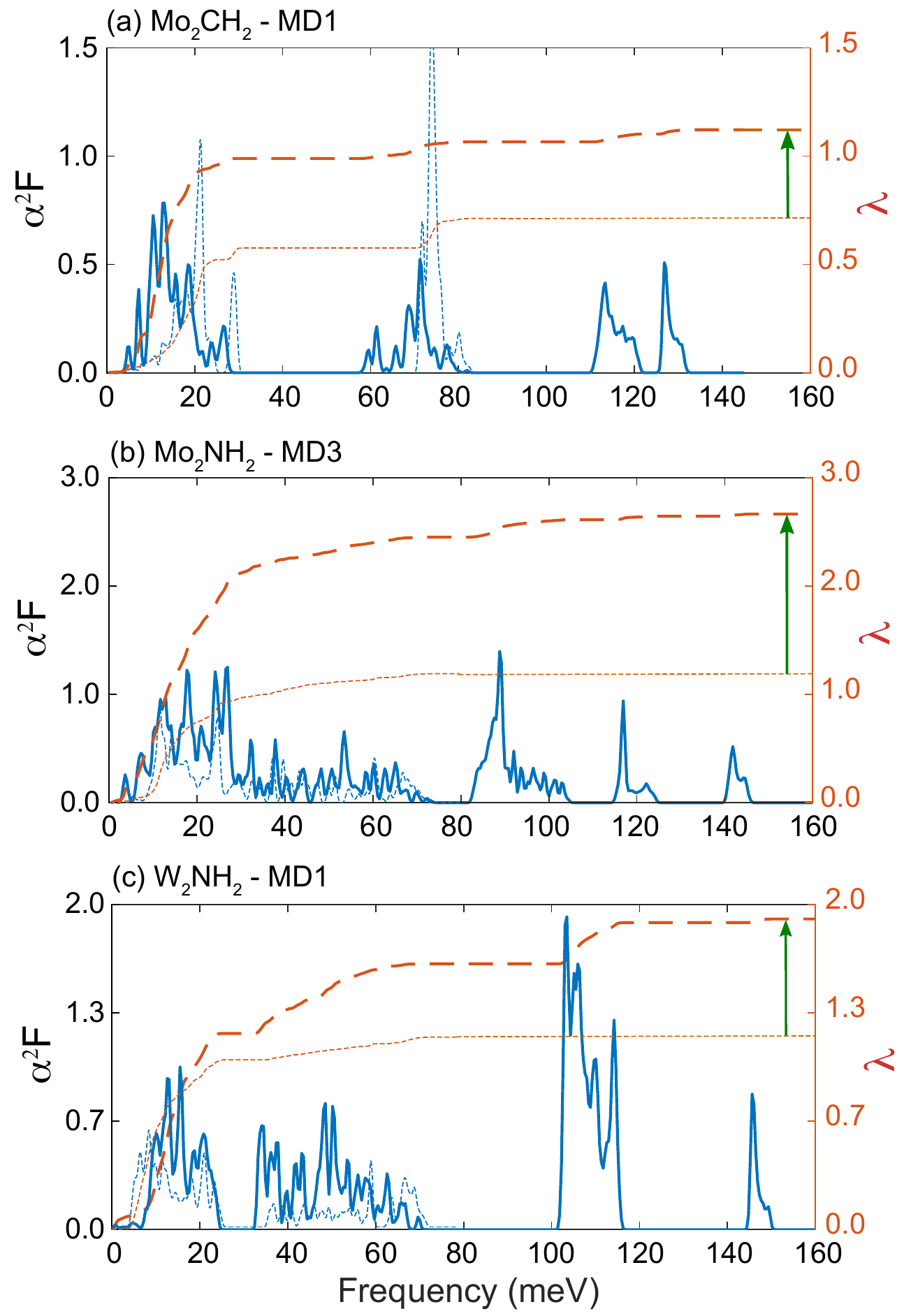}
\caption{\label{figure4} Electron-phonon coupling properties of hydrogenated MXenes with the highest $T_c$'s. Eliashberg function $\alpha^2F$ (Solid blue lines), and EPC function $\lambda$ (large-dashed red lines) of (a) Mo$_2$CH$_2$ (MD1), (b) Mo$_2$NH$_2$ (MD3), and (c) W$_2$NH$_2$ (MD1) as a function of phonon frequency. The small-dashed lines show the corresponding results for the pristine compounds. The green arrows indicate the enhancement of $\lambda$ due to hydrogenation.}
\end{figure}

\begin{table}[t]
\centering
\begin{tabular}{ ccccc }
\hline 
Compound & $N_F$ (eV$^{-1}$) & $v_F$ (10$^{6}$ ms$^{-1}$) & $T_c$ (K) & $\xi(0)$ (nm) \\ \hline 
Mo$_2$CH$_2$ -- MD1 & 2.09 & 0.311 & 12.6 & 25\\
Mo$_2$CH$_2$ -- MD3 & 2.50 & 0.409 & 11.8 & 35\\  
Mo$_2$CH$_2$ -- MD4 & 2.19 & 0.520 & 5.5 & 96\\ \hline
Mo$_2$NH$_2$ -- MD1 & 3.55 & 0.308 & 21.5 & 15\\
Mo$_2$NH$_2$ -- MD3 & 6.75 & 0.198 & 32.4 & 6\\
Mo$_2$NH$_2$ -- MD4 & 3.41 & 0.257 & 14.7 & 18\\ \hline
W$_2$NH$_2$ -- MD1 & 4.06 & 0.283 & 30.7 & 9\\
W$_2$NH$_2$ -- MD3 & 2.88 & 0.302 & 22.0 & 14\\
W$_2$NH$_2$ -- MD4 & 3.37 & 0.245 & 26.7 & 9\\
\hline 
\end{tabular}
\caption{\label{tab:table1}%
Calculated density of states at the Fermi level ($N_F$), averaged Fermi velocity ($v_F$), superconducting transition temperature ($T_c$), and coherence length at $T=0$ ($\xi(0)$) for the stable hydrogenated MXene structures. 
}
\end{table}

Subsequently, we have evaluated the Eliashberg spectral function and EPC constant from our DFPT results. We obtained considerable enhancements of the EPC constants upon hydrogenation for all considered compounds, as shown in Fig.~\ref{figure4}, albeit based on different mechanisms. The full results for the $T_c$'s of the different compounds are summarized in Table \ref{tab:table1}.

As discussed in Section \ref{sec:elect_vib}, the electronic DOS at $E_F$ of Mo$_2$CH$_2$ is depleted compared with pristine Mo$_2$C. This is in principle detrimental to the EPC since $\lambda \propto N_F$. However, this effect is amply compensated by the softening of the flexural ZA phonon mode, boosting $\lambda$ owing to its dependence on $\alpha^2F/\omega$ (as introduced in Section \ref{sec:methods}). Similar strong EPC hosted by a flexural mode also occurs in pristine Mo$_2$N, resulting in its enhanced superconducting $T_c$ compared with Mo$_2$C \cite{D0NR03875J}. As a result, the EPC constant is enhanced from $\lambda=0.75$ in pristine Mo$_2$C to $\lambda=1.12$ in Mo$_2$CH$_2$ (MD1), as shown in Fig.~\ref{figure4}(a), and the corresponding $T_c$ increases from 7.1 K to 12.6 K. MD of Mo$_2$CH$_2$ has a comparable $T_c$ of 11.8 K, while the $T_c$ of MD4 is significantly depleted (5.5 K). 

On the other hand, the main contribution to the enhanced $T_c$ of Mo$_2$NH$_2$ and W$_2$NH$_2$ is the increase in $N_F$ upon hydrogenation, as discussed in Section \ref{sec:elect_vib}. In the case of Mo$_2$NH$_2$ (MD3) (see Fig.~\ref{figure4}(b)), the EPC attains a particularly high value of 2.67 -- more than double the value obtained for pristine Mo$_2$N (1.2) \cite{D0NR03875J} -- yielding a high $T_c$ of 32.4 K. As MD1 and MD4 of  Mo$_2$NH$_2$ have a considerably lower DOS at $E_F$, their $T_c$'s are limited to 21.5 K and 14.7 K respectively -- the former still exceeding the $T_c$ of 16 K of pristine Mo$_2$N \cite{D0NR03875J}. 
 
Likewise, the EPC of W$_2$NH$_2$ (MD1) is enhanced to an elevated value of 1.91, yielding a $T_c$ of 30.7 K. Interestingly, the contribution of in-plane H modes, situated in the range 100--120 meV, to the EPC in W$_2$NH$_2$ is particularly strong, as shown in Fig.~\ref{figure4}(c). The other dynamically stable models of W$_2$NH$_2$ (MD3 and MD4) also produce sizeable $T_c$'s (22.0 K and 26.7 K). 

Our first-principles characterization of the different hydrogenated MXene compounds also enables the calculation of superconducting length scales, such as coherence lengths, through the Ginzburg-Landau relation $\xi(0)=\sqrt{\frac{7 \zeta(3)}{3}} \frac{\hbar v_F}{4 \pi T_c}$ \cite{Fetter}, where $v_F$ is the average Fermi velocity. The results are summarized in Table \ref{tab:table1}. Due to the elevated $T_c$ values of the hydrogenated MXenes, their coherence lengths are generally reduced compared to the pristine MXenes \cite{D0NR03875J}, with the notable exception of Mo$_2$CH$_2$ in MD4.

\section{Inducing superconductivity in niobium carbide}

\begin{figure}[h]
\includegraphics[width=\linewidth]{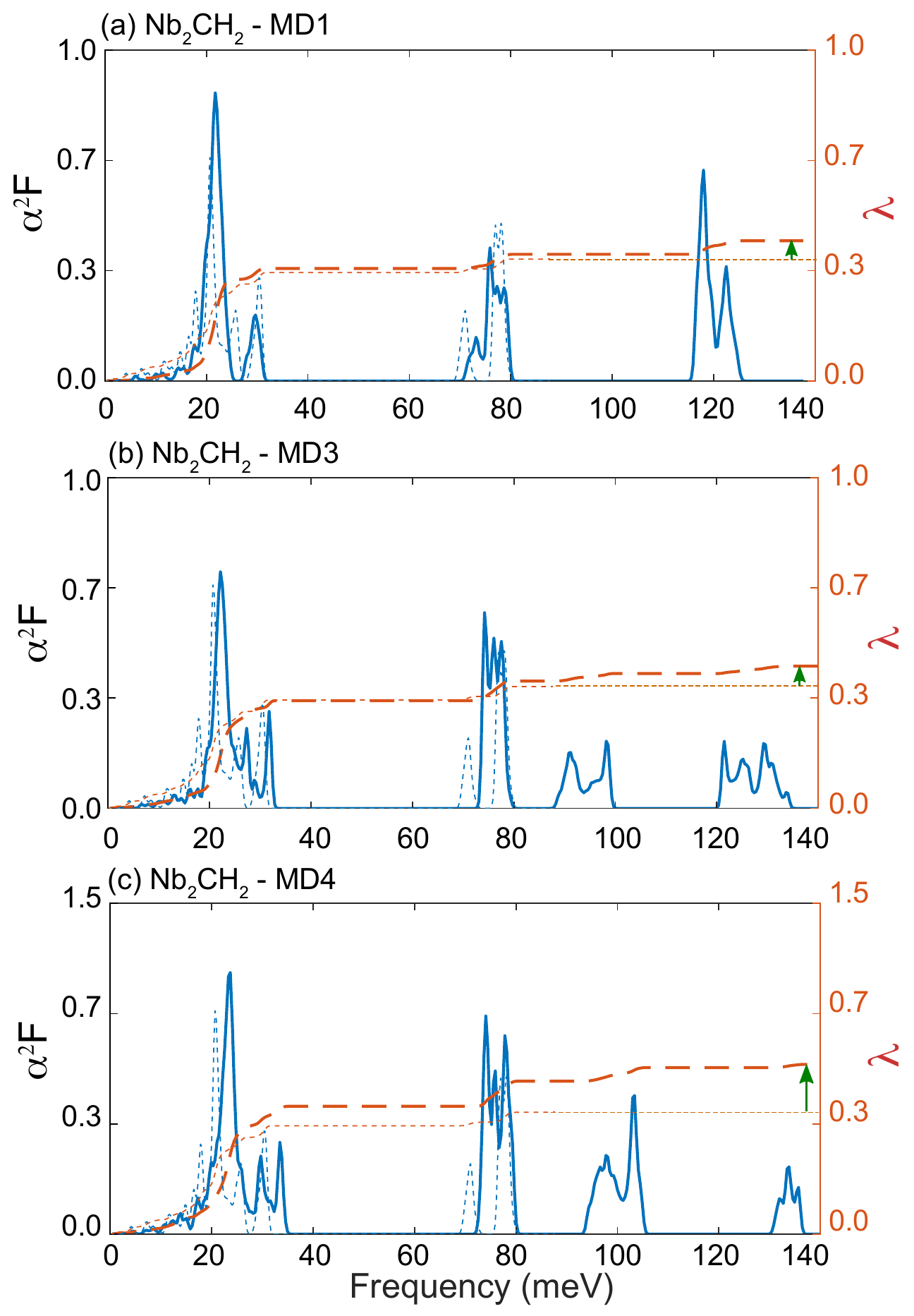}
\caption{\label{figure5} Electron-phonon coupling properties of Nb$_2$CH$_2$. Eliashberg function $\alpha^2F$ (solid blue lines), and EPC function $\lambda$ (large-dashed red lines) of (a) MD1, (b) MD3, and (c) MD4, as a function of phonon frequencies. The small-dashed lines show the corresponding results for the pristine compounds. The green arrows indicate the enhancement of $\lambda$ due to hydrogenation.}
\end{figure}

Earlier first-principles exploration of pristine MXenes identified one superconducting MXene based on a group-5 transition metal (V, Nb, Ta), namely Ta$_2$N with a $T_c$ of 2 K \cite{D0NR03875J}. However, the majority of MXene compounds based on group-5 elements were found to be non-superconducting, due to a lack of sufficiently strong EPC. The absence of superconductivity in pristine Nb$_2$C has indeed been confirmed experimentally \cite{Kamysbayev979}. While previous experimental studies have shown the emergence of superconductivity in Nb$_2$C under the influence of surface functionalization \cite{Kamysbayev979,Babar_2020,WANG2022101711}, reaching $T_c$’s up to 6 K in case of Cl and S adatoms, and up to 7 K for NH functional groups \cite{Kamysbayev979}, the effect of hydrogen adatoms has not yet been considered.

Just as for the Mo- and W-based hydrogenated MXenes presented above, MD2 of Nb$_2$CH$_2$ is found to be dynamically unstable, having evanescent phonon modes with predominant hydrogen character (see the SM). In case of MD1 and MD3, the contributions of niobium and carbon to the Eliashberg function and the EPC are largely unaltered compared with pristine Nb$_2$C, as shown in Fig.~\ref{figure5}(a,b). However, the hydrogen-based phonon modes contribute additional EPC channels, enhancing the EPC constant $\lambda$ to 0.42 and 0.43 for MD1 and MD3 respectively. A notable difference between these two cases is that the hydrogen phonon modes form a single band in MD1 (centered around 120 meV), while in MD3 they split into two separate bands (the lower band lies in the range of $90-100$ meV and the upper band in the range of $120-135$ meV; see also the SM). The relatively modest enhancements of the EPC due to the hydrogen phonon modes nevertheless suffice to induce superconductivity in MD1 and MD3 of Nb$_2$CH$_2$, with $T_c$'s of 0.8 K and 1.0 K respectively. 

As shown in Fig.~\ref{figure5}(c), MD4 of Nb$_2$CH$_2$ hosts enhanced coupling to niobium- and carbon-based phonon modes (as opposed to MD1 and MD3), combined with additional EPC due to hydrogen modes, yielding a total EPC constant $\lambda$ of 0.51 and a $T_c$ of 2.9 K. 

\section{\label{sec:Concl}Conclusions}

Our first-principles calculations on hydrogenated MXenes have revealed consistently improved superconducting properties compared to their pristine counterparts (described in Ref.~\citenum{D0NR03875J}). Our analysis of the dynamical and thermodynamic stability of hydrogenated molybdenum- and tungsten-based MXenes has yielded three different stable structures (see Fig.~\ref{figure1}), consistently harboring enhanced EPC and therefore elevated superconducting $T_c$'s.  

Several mechanisms for the enhanced superconducting properties were identified. Firstly, enhanced EPC due to a softened flexural phonon mode, e.g.~in the Mo-based flexural mode of Mo$_2$CH$_2$. Secondly, an increase in the electronic DOS at the Fermi level due to charge transfer from hydrogen to the MXene layer occurs in several compounds, such as Mo$_2$NH$_2$ and W$_2$NH$_2$. Combined with flat segments in the electronic dispersion, this results in a Van Hove singularity in the DOS around the Fermi level in Mo$_2$NH$_2$ (see Fig.~\ref{figure2}(d)) -- akin to the case of hydrogenated monolayer MgB$_2$ \cite{PhysRevLett.123.077001}. This enhances the EPC constant to 2.67, more than double the value obtained for pristine Mo$_2$N \cite{D0NR03875J}, and an elevated $T_c$ of 32 K. Finally, additional contributions to the EPC appear upon hydrogenation, directly stemming from the hydrogen-based phonon modes. The latter mechanism is observed in all studied compounds, but is particularly strong in case of the in-plane hydrogen phonon modes of W$_2$NH$_2$ (see Fig.~\ref{figure4}(c)). 

In addition, we have demonstrated that superconductivity can even be induced by hydrogenation in MXenes which are otherwise non-superconducting in pristine form, such as Nb$_2$C (see Fig.~\ref{figure5}). 

Altogether our results clearly demonstrate that the superconducting properties of the broad family of MXene materials can be engineered through surface functionalization, corroborating recent experimental advances \cite{Kamysbayev979,Babar_2020,WANG2022101711}. Moreover, the hydrogenated MXenes studied here present another important class of 2D materials in which hydrogen adatoms boost the EPC and the superconducting $T_c$, extending significantly prior results for \textit{p}-doped graphane \cite{PhysRevLett.105.037002}, hydrogenated monolayer MgB$_2$ \cite{PhysRevLett.123.077001}, and hydrogenated gallium monolayers (gallenane) \cite{Petrov_2021}. 

\section*{Conflicts of interest}
There are no conflicts to declare.

\section*{Acknowledgements}
This work is supported by Research Foundation-Flanders (FWO), The Scientific and Technological Research Council of Turkey (TUBITAK) under the contract number COST-118F187, the Air Force Office of Scientific Research under award number FA9550-19-1-7048. Computational resources were provided by the High Performance and Grid Computing Center (TRGrid e-Infrastructure) of TUBITAK ULAKBIM and by the VSC (Flemish Supercomputer Center), funded by the FWO and the Flemish Government -- department EWI. J.B. is a postdoctoral fellow of the FWO.



\bibliography{References} 
\bibliographystyle{rsc} 

\end{document}